\documentclass[pra,preprint,letterpaper]{revtex4}

\usepackage{graphicx}

\usepackage{amsthm}

\usepackage{amsfonts}

\begin{document}

\bibliographystyle{apsrev}

\newcommand{\Bra}[1]{\ensuremath{\left \langle #1 \right |}}
\newcommand{\Ket}[1]{\ensuremath{\left | #1 \right \rangle}}
\newcommand{\tr}{\ensuremath{\mbox{Tr}}}
\renewcommand{\H}{\ensuremath{\mathcal{H}}}
\newcommand{\E}{\ensuremath{\mathcal{E}}}

\title{Optimal Entanglement Generation from Quantum Operations}
\author{M. S. Leifer}
\email[Email: ]{Matt.Leifer@bristol.ac.uk}
\author{L. Henderson}
\author{N. Linden}
\affiliation{Dept. of Mathematics, University of Bristol, University Walk, Bristol, BS8 1TW, UNITED KINGDOM}
\date{\today}

\begin{abstract}
We consider how much entanglement can be produced by a non-local two-qubit unitary operation, $U_{AB}$ - the \emph{entangling capacity} of $U_{AB}$.  For a single application of $U_{AB}$, with no ancillas, we find the entangling capacity and show that it generally helps to act with $U_{AB}$ on an entangled state.  Allowing ancillas, we present numerical results from which we can conclude, quite generally, that allowing initial entanglement typically increases the optimal capacity in this case as well.  Next, we show that allowing collective processing does not increase the entangling capacity if initial entanglement is allowed.
\end{abstract}

\maketitle

\section{Introduction}

The fundamental resource used in many quantum information protocols, such as cryptography and teleportation, is the entanglement in a quantum state.  A major theme of investigation in quantum information theory is the analysis and characterisation of entanglement properties of quantum states under local operations and classical communication (LOCC).  One issue is how to extract the entanglement in a quantum state.  The simplest protocols involve taking a single copy of the quantum state and using LOCC to extract the entanglement \cite{NeilMon}.  An important realisation is that, in general, collective processing (i.e. processing more than one copy of the state at a time) is more efficient than individual copy processing.  Indeed, for mixed states \cite{PurColl}, there are examples where no entanglement can be extracted at all if one only has one copy, but collective processing does allow extraction of entanglement.  The fact that \emph{asymptotic} collective processing (i.e. processing of infinitely many copies) is necessary for the \emph{reversible} extraction of entanglement is a key building block in the general theory of entanglement \cite{Concentrate, Reverse}.

The fundamental resource used in quantum control theory and quantum computing is a non-local quantum operation, such as an interaction Hamiltonian or a unitary gate.  These can be used, along with local actions, to perform the steps of quantum algorithms and generate entangled states.  Conversely, an entangled state and LOCC can be used to apply a non-local operation to an arbitrary state, enabling distributed quantum processing.

Just as for quantum states, it is important to find ways of classifying and quantifying the non-local properties of operations. There is a multitude of inter-related problems here.  Indeed, there seems to be an even richer structure in the case of quantum operations than there is for states.  For example, one can consider how much entanglement an operation can generate, how much classical communication the operation can perform, or the power of the operation to simulate other operations.  As with states, we may restrict ourselves to a single application of an operation or we may process multiple copies collectively.

This area has attracted much interest recently and results have been obtained on Hamiltonian simulation \cite{2qSim, UniSim, NeilHam1, NeilHam2, Beth1, Beth2, Beth3, Beth4, Leung1, ChenSim, CiracSim1, CiracSim2, CiracSim3}, interconversion of unitary operations \cite{CiracCon1, CiracCon2}, entanglement generation \cite{Hamil1, CiracDecomp, Zanardi1, Zanardi2} and generating operations from entangled states \cite{CiracStatOp, NonOp, PlenOp}.  Most of these results have focussed on protocols involving a single application of the operation and little is known about the multi-copy and asymptotic cases.

In this paper, we focus on the problem of entanglement generation for two-qubit unitary operations acting on pure states.  Suppose that Alice and Bob share a state $\Ket{\psi}$ in their combined Hilbert space $\H_A \otimes \H_B$ and that they are able to implement an operation $U_{AB} \in U(4)$ on any non-local two-qubit subspace.   They would like to maximise the amount of entanglement that they generate per application of $U_{AB}$.  We call this maximum the entangling capacity, $\mathcal{EC}_E$, of $U_{AB}$.  For single applications of $U_{AB}$, the entangling capacity is given by
\begin{equation}
\mathcal{EC}_E\left (U_{AB}\right ) = \mbox{max}_{\Ket{\psi} \in \H_A \otimes \H_B} \left [ E\left ( U_{AB}\Ket{\psi} \right ) - E\left (\Ket{\psi} \right )\right ]
\end{equation}
where $E$ is an entanglement measure and $U_{AB}$ acts on one qubit in $\H_A$ and one in $\H_B$.

In \S\ref{Kraus}, we review the useful decomposition of two-qubit unitaries that was introduced in \cite{CiracDecomp,UniSim}.  \S\ref{single} of the paper concerns the single-copy entangling capacity.  In \S\ref{purity}, we review an argument due to \cite{IBMent, IBMClass} that shows that the single-copy entangling capacity can be achieved when $U_{AB}$ is only allowed to act on pure states.  We then extend this argument to show that pure states can still be used if the entangling capacity is to be achieved using the minimal amount of initial entanglement.    In \S\ref{NoAnc} and \S\ref{Anc}, we show how much entanglement can be created by a single use of a quantum operation when we allow Alice and Bob to share initial entanglement; this work extends \cite{CiracDecomp} where the authors considered entangling capacities of unitaries but did not allow initial entanglement; it also extends \cite{Hamil1}, which allowed initial entanglement but only unitary transformations infinitesimally close to the identity (i.e. Hamiltonians).  In the case where ancillas are not allowed (\S\ref{NoAnc}), we are able to derive analytic results about the entangling capacities of unitaries.  We find that it generally helps to start with an entangled state, although this is dependent on the entanglement measure.  \S\ref{Anc} concerns the case where we allow ancillas; we mostly describe numerical results here, however these numerical results allow us to conclude, quite generally, that allowing initial entanglement can increase the entangling capacity even when ancillas are available.

The final part of this paper (\S\ref{Coll}) concerns collective processing of quantum operations.  As described above, collective processing is a key idea in understanding entanglement properties of quantum states.  Our main result, essentially that collective processing of quantum operations does not help in generating quantum entanglement, is in stark contrast to the situation for processing of quantum states.  We conclude with a discussion of the implications of these results for the  interconvertibility of quantum operations and the classification of their entanglement properties.

\section{Decomposition of two-qubit unitary operators}
\label{Kraus}

The entanglement properties of a unitary operation are invariant under local unitary operations applied before or after the operation.  This gives a notion of local equivalence of operations
\begin{equation}
U_{AB} \sim U'_{AB} \,\, \mbox{iff} \,\, U'_{AB} = V_A \otimes V_B U_{AB} W_A \otimes W_B
\end{equation}
where $V_A, V_B, W_A, W_B$ are local unitaries acting on the systems indicated. In order to simplify our calculations, we make use of the following decomposition of two-qubit unitary operators.  Any two-qubit unitary, $U_{AB} \sim U_d$, where
\begin{equation}
\label{Kraus:Ud}
U_d = e^{i \sum_{j=1}^3 \alpha_j \sigma^A_j \otimes \sigma^B_j}
\end{equation}
$\frac{\pi}{4} \geq \alpha_1 \geq \alpha_2 \geq |\alpha_3| \geq 0$ and $\sigma_{1,2,3}$ are the Pauli matrices.  Since, $U_d$ has the same entangling capacity as $U$, we always work with this form \footnote{In fact, when considering the entangling capacity, we can always take $\alpha_3 \geq 0$.  This is because $e^{i \sum_{j=1}^2 \alpha_j \sigma^A_j \otimes \sigma^B_j - \alpha_3 \sigma^A_3 \otimes \sigma^B_3} \sim \left ( e^{i \sum_{j=1}^3 \alpha_j \sigma^A_j \otimes \sigma^B_j} \right ) ^*$ and $\mathcal{EC}_E$ is invariant under conjugation}.  Note that the eigenvalues of $U_d$ are given by $e^{i\lambda_j}$ where
\begin{equation}
\label{Kraus:Eigen}
\begin{array}{llllllll}
\lambda_1 & = & - & \alpha_1 & + & \alpha_2 & + & \alpha_3 \\
\lambda_2 & = & + & \alpha_1 & - & \alpha_2 & + & \alpha_3 \\
\lambda_3 & = & + & \alpha_1 & + & \alpha_2 & - & \alpha_3 \\
\lambda_4 & = & - & \alpha_1 & - & \alpha_2 & - & \alpha_3
\end{array}
\end{equation}
The corresponding eigen-basis is given by $U_d\Ket{\Phi_j} = e^{i\lambda_j}\Ket{\Phi_j}$ and is the Bell basis.  For later convenience, we choose the following phase convention:
\begin{equation}
\label{Kraus:Evect}
\begin{array}{lll}
\Ket{\Phi_1} & = & \frac{-i}{\sqrt{2}} \left( \Ket{00} - \Ket{11} \right) \\
\Ket{\Phi_2} & = & \frac{1}{\sqrt{2}} \left( \Ket{00} + \Ket{11} \right) \\
\Ket{\Phi_3} & = & \frac{-i}{\sqrt{2}} \left( \Ket{01} + \Ket{10} \right) \\
\Ket{\Phi_4} & = & \frac{1}{\sqrt{2}} \left( \Ket{01} - \Ket{10} \right)
\end{array}
\end{equation}
In \cite{CiracDecomp}, an explicit method is given for calculating $\alpha_j, V_A, V_B, W_A$ and $W_B$ for any unitary.  However, since we are only be interested in the values $\alpha_j$, the following method can be used.

Firstly, define
\begin{equation}
\tilde{U} = \sigma_2 \otimes \sigma_2 U^T \sigma_2 \otimes \sigma_2
\end{equation}
where $^T$ indicates the transpose in the computational basis.  The eigenvalues of $\tilde{U}U$ are local invariants of $U$, equivalent to those found in \cite{Makhlin}.  From, eq.(\ref{Kraus:Ud}) one can see that these invariants are in fact squares of the eigenvalues of $U_d$.  Thus, solving eq.(\ref{Kraus:Eigen}) gives the unique decomposition.

\section{Single Copy Entangling Capacity}

\label{single}

\subsection{Purity of States in the Optimal Protocol}

\label{purity}

In this section we determine whether optimal protocols can be found for generating entanglement using one application of $U_{AB}$ that only involve pure states at every stage.  We use an argument of \cite{IBMent, IBMClass} to establish that this is the case. Further, we extend this argument to show that optimal pure state protocols can be found that start with the minimum possible amount of initial entanglement.  Thus, all the important details of the single-copy entangling capacity of $U_{AB}$ can be established by considering pure states only.

Making a suitable definition of the entangling capacity over 
mixed states is not quite as straightforward as the pure state 
case.  In particular, the choice of entanglement measure for the 
initial and final states may be different.  For the initial 
state, it seems natural to use a measure of the minimum average amount of 
entanglement required to generate it (i.e. the entanglement of formation).  However, for the final 
state it makes more sense to measure the maximum amount of 
entanglement that can be extracted from it (i.e. the distillable entanglement).  

To make this more specific, consider an initial mixed state 
$\rho_0$.  Let $\rho_0 = \sum_j p_j \Ket{\psi_j}\Bra{\psi_j}$ be 
the decomposition of $\rho_0$ with minimal ensemble average 
entanglement.  To generate an ensemble of $n$ states described by 
$\rho_0$, we may prepare $\Ket{\psi_j}$ with probability $p_j$ and 
then discard the information about which state was prepared.  As 
$n \rightarrow \infty$ , the amount of entanglement per state 
used in this procedure will be $E_f(\rho_0)$, where $E_f$ is the 
entanglement of formation.  The operation $U_{AB}$ can then be 
applied to each state individually yielding $n$ copies of the 
state $\rho_1 = U_{AB}\rho_0U_{AB}^\dagger$.  These states can 
then be distilled to singlets by LOCC and as $n \rightarrow 
\infty$ the yield of singlets per copy of $\rho_1$ will be 
$D(\rho_1)$, where $D$ is the distillable entanglement.  Note 
that, although this protocol involves collective processing of 
the states, the fact that $U_{AB}$ is applied to each copy of 
$\rho_0$ individually means that it can still be regarded as a 
single-copy protocol with respect to the non-local operation.

With this in mind, we define the mixed state single-copy 
entangling capacity, $C^{mixed}_E$ as 
\begin{equation} 
C^{mixed}_E = \mbox{max}_{\rho_0}(D(\rho_1) - E_f(\rho_0)) 
\end{equation} 
Then 
\begin{eqnarray} D(\rho_1) - E_f(\rho_0) & 
\leq & E_f(\rho_1) - E_f(\rho_0) \nonumber \\
              & \leq & \sum_j p_j \left ( E_f(U_{AB}\Ket{\psi_j}\Bra{\psi_j}U^{\dagger}_{AB}) - E_f(\Ket{\psi_j}\Bra{\psi_j})\right )               \label{purity:popt1} \\
              & \leq & \mbox{max}_{\psi_j} \left ( E_f(U_{AB}\Ket{\psi_j}) - E_f(\Ket{\psi_j}) \right ) \label{purity:popt2}
\end{eqnarray}
This demonstrates that for every mixed state, there is a pure state for which the action of $U_{AB}$ generates at least as much entanglement.

Next we show that any mixed state that achieves the entangling capacity cannot be formed using less entanglement than there is in a pure state that achieves the entangling capacity with minimal initial entanglement.  Let $\Ket{\psi}$ be a pure state that achieves the entangling capacity with the minimal possible initial entanglement.  Let $\rho$ be a mixed state that also achieves the entangling capacity.  From eqs.(\ref{purity:popt1}) and (\ref{purity:popt2}) it is clear that the optimal decomposition of $\rho$ must be a mixture of pure states that achieve the entangling capacity.  Since this is the optimal decomposition of $\rho$, $E_f(\rho)$ is just the weighted average of the entanglements of these pure states.  Thus, $E_f(\rho) \geq E_f(\Ket{\psi})$ because $\Ket{\psi}$ has the minimal entanglement of any possible state in this ensemble.

\subsection{Single Application with no ancillas}
\label{NoAnc}

We now determine the entangling capacity of two-qubit unitaries of the form of eq.(\ref{Kraus:Ud}) when no ancillas are allowed.  This depends on the entanglement measure we choose to optimise over.  In \S\ref{sqcon} we optimise over the square of concurrence and then in \S\ref{other} we show how our results can be extended to other measures of entanglement.




\subsubsection{Square of concurrence}
\label{sqcon}
One entanglement measure that is particularly convenient to optimise is the square of the concurrence \cite{WootCon}, $C$, defined by
\begin{equation}
C(\Ket{\psi}) = \left | \Bra{\psi} \sigma_2 \otimes \sigma_2 \Ket{\psi^*} \right |
\end{equation}
where $\Ket{\psi^*}$ is the state vector obtained by taking the complex conjugates of the components of $\Ket{\psi}$ in the computational basis.  We can adapt an argument in \cite{CiracDecomp} to perform the optimisation here.

Writing $\Ket{\psi} = \sum_j b_j \Ket{\Phi_j}$ gives
\begin{equation}
\label{single:delta}
\Delta C^2 = C_f^2 - C_0^2 = \left | \sum_j e^{2i\lambda_j} b_j^2 \right | ^2 - \left | \sum_j b_j^2 \right | ^2 = \sum_{j,k} \left ( e^{2i(\lambda_j - \lambda_k)} - 1\right ) b_j^2 b_k^{*2}
\end{equation}
where $C_0$ is the initial concurrence and $C_f$ is the final concurrence after applying $U_{AB}$.

This can be optimised by imposing the normalisation condition $\sum_j \left | b_j\right |^2 = 1$ with a Lagrange multiplier, $2\mu$, i.e. we maximise
\begin{equation}
L = \sum_{j,k} \left ( e^{2i(\lambda_j - \lambda_k)} - 1\right ) b_j^2 b_k^{*2} - 2\mu \left ( \sum_j b_j b_j^* - 1\right )
\end{equation}
Differentiating gives
\begin{equation}
\label{single:diff}
\frac{\partial L}{\partial b_j} = 2b_j e^{2i\lambda_j} \sum_k e^{-2i\lambda_k} b_k^{*2} - 2b_j \sum_k b_k^{*2} - 2\mu b_j^* = 0
\end{equation}
multiplying by $b_j$ and summing over $j$ gives
\begin{equation}
\sum_{j,k} \left ( e^{2i(\lambda_j - \lambda_k)} - 1\right )b_j^2 b_k^{*2} - \mu \sum_j \left | b_j \right | ^2 = 0
\end{equation}
which yields
\begin{equation}
\label{single:multsolv}
\mu = C_f^2 - C_0^2
\end{equation}
Substituting eqs.(\ref{single:multsolv}) and (\ref{single:delta}) into eq.(\ref{single:diff}) gives
\begin{equation}
\label{single:quad}
b_j e^{2i\lambda_j} e^{2i\eta} C_f - b_j e^{2i\epsilon} C_0 - C_f^2 b_j^* + C_0^2 b_j^* = 0
\end{equation}
where $\epsilon, \eta$ are phases depending on all of the $b_j$'s.
One possible solution is $b_j = 0$.  To find the other solutions we write $b_j = \beta_j e^{i\gamma_j}$ where $\beta_j, \gamma_j \in \mathbb{R}$.  These solutions must have $\beta_j \neq 0$ and so eq.(\ref{single:quad}) reduces to
\begin{equation}
\label{single:quad2}
C_f^2 - e^{2i(\lambda_j + \gamma_j + \eta)}C_f - C_0^2 + e^{2i(\gamma_j + \epsilon)} C_0 = 0
\end{equation}
There are as many equations (\ref{single:quad2}) as there are non-zero $b_j$'s.  For generic $\lambda_j$'s, we will show that at most two of these equations can be satisfied simultaneously.

First, consider the case when the optimal starting state has $C_0 = 0$.  Then we have
\begin{equation}
C_f \left ( C_f - e^{2i (\lambda_j + \gamma_j + \eta)} \right ) = 0
\end{equation}
Since $C_f$ is real and we are looking for the maximum, we must have $C_f = 1$.  This shows that it is only best to start in a product state if $U_{AB}$ can generate one e-bit of entanglement when no ancillas are present.  The conditions for this were found in \cite{CiracDecomp} to be
\begin{equation}
\label{single:cond}
\alpha_1 + \alpha_2 \geq \frac{\pi}{4} \,\, \mbox{and} \,\,
\alpha_2 + \alpha_3 \leq \frac{\pi}{4}
\end{equation}
so here we will focus on the cases where (\ref{single:cond}) is violated and the optimal starting state must have non-zero $C_0$.

Subtracting any two of eqs.(\ref{single:quad2}) gives
\begin{equation}
\label{single:subtract}
\sin \left ( \lambda_j - \lambda_k + \gamma_j - \gamma_k \right ) C_f = e^{i (2\epsilon - 2\eta - \lambda_j - \lambda_k)} \sin \left ( \gamma_j - \gamma_k \right )C_0
\end{equation}
This gives consistency conditions for the simultaneous solution of any pair of eqs.(\ref{single:quad2}).  In particular, since $C_f$ and $C_0$ are both real, we have that
\begin{equation}
\label{single:consist}
2 \left ( \epsilon - \eta \right ) - \lambda_j -\lambda_k = n \pi, n \in \mathbb{Z}
\end{equation}
For generic $\lambda_j$'s this condition cannot be satisfied for more than one pair of equations in (\ref{single:quad2}).  Thus, at most two $b_j$'s can be non-zero \footnote{This result can be extended to all possible $\lambda_j$'s by noting that eq.(\ref{single:subtract}) can only be satisfied for more than one pair if some of the eigenvalues are degenerate.  Further, it can be shown that one can choose only one of the corresponding $b_j$'s to be non-zero.}.  This means that the optimal starting state will always be in a subspace spanned by two of the eigenvectors of $U_{AB}$.  We will choose the two eigenvectors and the coefficients $b_j$ that maximise $\Delta C^2$.  Re-expressing eq.(\ref{single:delta}) in terms of $\beta_j, \gamma_j$ gives
\begin{equation}
\Delta C^2 = 4 \sum_{j<k} \beta_j^2 \beta_k^2 \left [ \sin \left ( 2 \left ( \gamma_j - \gamma_k \right ) + \lambda_j - \lambda_k \right ) \sin \left( \lambda_k - \lambda_j \right ) \right ]
\end{equation}
Only one term in this sum can be non-zero and for this term we may choose $\gamma_j, \gamma_k$ so that $\Delta C^2 = 4\beta_j^2 \beta_k^2 \left | \sin (\lambda_k - \lambda_j) \right |$.  This is maximised by $\beta_j = \beta_k = \frac{1}{\sqrt{2}}$.  Thus the entangling capacity is given by
\begin{equation}
\mathcal{EC}_{C^2} = \mbox{max}_{j < k} \left | \sin(\lambda_k - \lambda_j) \right |
\end{equation}
Note that this is greater than the corresponding result of $\mbox{max}_{j < k} \left | \sin(\lambda_k - \lambda_j) \right |^2$ found in \cite{CiracDecomp} when the starting state is restricted to be a product.  This shows that when (\ref{single:cond}) is violated, initial entanglement is always required to achieve the optimal capacity when no ancillas are allowed.  There are two parameter regions where (\ref{single:cond}) does not hold.
\begin{enumerate}

\item $\alpha_1 + \alpha_2 < \frac{\pi}{4}, \alpha_2 + \alpha_3 < \frac{\pi}{4}$.  In this region, the maximum is given by making the $j = 3, k = 4$ term nonzero.  We find that $\mathcal{EC}_{C^2} = \sin(2(\alpha_1 + \alpha_2))$ and the optimal starting state is $\Ket{\psi} = \left ( \sin(\frac{\alpha_1 + \alpha_2}{2} - \frac{\pi}{8})\Ket{01} -i \cos(\frac{\alpha_1 + \alpha_2}{2} - \frac{\pi}{8})\Ket{10} \right )$.  This gives an optimal initial entanglement of $C_0^2 = \frac{1}{2} \left ( 1 - \sin2(\alpha_1 + \alpha_2)\right )$

\item $\alpha_1 + \alpha_2 > \frac{\pi}{4}, \alpha_2 + \alpha_3 > \frac{\pi}{4}$.  In this region, the maximum is given by making the $j = 1, k = 4$ term nonzero.  We find that $\mathcal{EC}_{C^2} = \sin(2(\alpha_2 + \alpha_3))$ and the optimal starting state is $\Ket{\psi} = \frac{1}{\sqrt{2}} \left ( \Ket{\Phi_1} + e^{i(\frac{\pi}{4} + \alpha_2 + \alpha_3)} \Ket{\Phi_4} \right )$.

\end{enumerate}
Note that the entangling capacity is always found to be a function of $\alpha_1 + \alpha_2$ or $\alpha_2 + \alpha_3$, i.e. a sum of two of only two of the parameters of the unitary.  The value of the third parameter does not affect the entangling capacity at all when no ancillas are allowed.

\subsubsection{Other entanglement measures}
\label{other}
All bipartite entanglement measures, $E$, are monotonic functions of one another and in particular of the concurrence squared (i.e. $E = E(C^2)$).  Generalising the strategy of eqs. (\ref{single:delta}-\ref{single:subtract}) to an arbitrary entanglement measure, $E$, by making use of $\frac{\partial E}{\partial b_j} = \frac{\partial E}{\partial (C^2)}\frac{\partial (C^2)}{\partial b_j}$ gives
\begin{equation}
\label{single:othermon}
\sin \left ( \lambda_j - \lambda_k + \gamma_j - \gamma_k \right ) C_f \frac{d E_f}{d (C_f^2)} = e^{i (2\epsilon - 2\eta - \lambda_j - \lambda_k)} \sin \left ( \gamma_j - \gamma_k \right )C_0 \frac{d E}{d (C_0^2)}
\end{equation}
This gives the same consistency conditions as eq.(\ref{single:consist}) so we still have that at most two $b_j$'s can be non-zero.  The only exception is when $\frac{d E}{d (C^2)} \propto \frac{1}{C}$, which occurs when our entanglement measure is the concurrence itself.  In this case similar methods show that the only consistent solutions are $C_0 = 0$ and $C_0 = 1$ meaning that the optimal starting state must always be a product.

\begin{figure}
\setlength{\unitlength}{7cm}
\begin{picture}(1,0.85)
\put(0,0){\includegraphics[angle = 0, width = 7.0cm, height = 6.0cm]{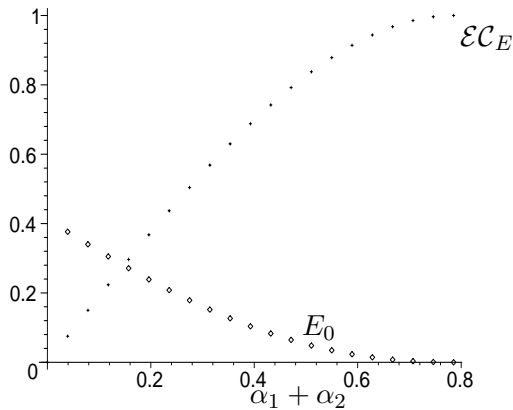}}
\put(0.5,0.02){$\alpha_1 + \alpha_2$}
\put(0.9,0.70){$\mathcal{EC}_{E}$}
\put(0.6,0.15){$E_0$}
\end{picture}
\caption{\label{graph:noanc}Single-copy entangling capacity and optimal initial entanglement for a general two-qubit unitary of the form of eq.(\ref{Kraus:Ud}) when no ancillas are allowed.  Crosses show the entangling capacity and diamonds show the minimum initial entanglement of a state that achieves the capacity.}
\end{figure}

For all other entanglement measures we focus on the case where $\alpha_1 + \alpha_2 < \frac{\pi}{4}, \alpha_2 + \alpha_3 < \frac{\pi}{4}$.  If we choose only $b_j$ and $b_k$ to be non-zero for some choice of $j \neq k = 1,2,3,4$ then the resulting optimal $\Delta E$ is always a function of the corresponding $\lambda_j$ and $\lambda_k$ only.  In fact, it must be the same function of $\lambda_j$ and $\lambda_k$ for all choices of $j$ and $k$.  For all the measures considered below we found that the optimal $\Delta E$ is always a monotonically increasing function of $|\lambda_j - \lambda_k|$ \footnote{Similarly when $\alpha_1 + \alpha_2 > \frac{\pi}{4}, \alpha_2 + \alpha_3 > \frac{\pi}{4}$ we found that, for any choice of $j$ and $k$, the optimal $\Delta E$ is always a monotonically decreasing function of $|\lambda_j - \lambda_k |$.}.   As with the square of concurrence, we choose the $j$ and $k$ that give the largest value of $|\lambda_j - \lambda_k|$, namely $j=3, k = 4$.  Thus, we can write the optimal starting state in its Schmidt decomposition as
\begin{equation}
\Ket{\psi} = \cos(\theta)\Ket{01} + e^{i \phi} \sin(\theta)\Ket{10}
\end{equation}
and we simply have to optimise $\Delta E$ over the Schmidt parameter $\theta$ and relative phase $\phi$.  We found the following results.
\begin{enumerate}

\item Concurrence:  $C = |\Bra{\psi}\sigma_2\otimes\sigma_2\Ket{\psi^*}|$.  As discussed above, this measure is unusual in that we must always start from a product state.   Thus, $\mathcal{EC}_{C} = \sin(2(\alpha_1 + \alpha_2))$, which coincides with the result of \cite{CiracDecomp}.

\item Entropy of entanglement: $E = -\tr(\rho^A \log_2 \rho^A)$, where $\rho^A$ is Alice's reduced density matrix.  We end up with a transcendental equation in $\theta$, which can be optimised numerically for each $\alpha_1 + \alpha_2$.  For results see Fig. \ref{graph:noanc}.

\item Linearised entropy: $R = 1 - \tr\left ( \left (\rho^A \right )^2 \right )$.  We find that $\mathcal{EC}_{R} = \sin(2(\alpha_1 + \alpha_2))$.

\end{enumerate}



\subsection{Ancillas}

Next we consider whether adding ancillas can increase the entangling capacity.  We have not yet solved this problem analytically, but we present some numerical optimisations, using entropy of entanglement as the measure.  We chose three different families of operations:
\begin{itemize}
\item The CNOT family $e^{i \alpha \sigma^A_1 \otimes \sigma^B_1}$.
\item The double CNOT (DCNOT) family $e^{i \alpha \left ( \sigma^A_1 \otimes \sigma^B_1 + \sigma^A_2 \otimes \sigma^B_2\right )}$.
\item The SWAP family $e^{i \alpha \left ( \sigma^A_1 \otimes \sigma^B_1 + \sigma^A_2 \otimes \sigma^B_2 + \sigma^A_3 \otimes \sigma^B_3\right )}$.
\end{itemize}
The families are so named because setting $\alpha = \frac{\pi}{4}$ gives operations that are locally equivalent to the CNOT, DCNOT and SWAP operations.

\begin{figure}
\setlength{\unitlength}{7cm}
\begin{picture}(1,0.85)
\put(0,0){\includegraphics[angle = 0, width = 7.0cm, height = 6.0cm]{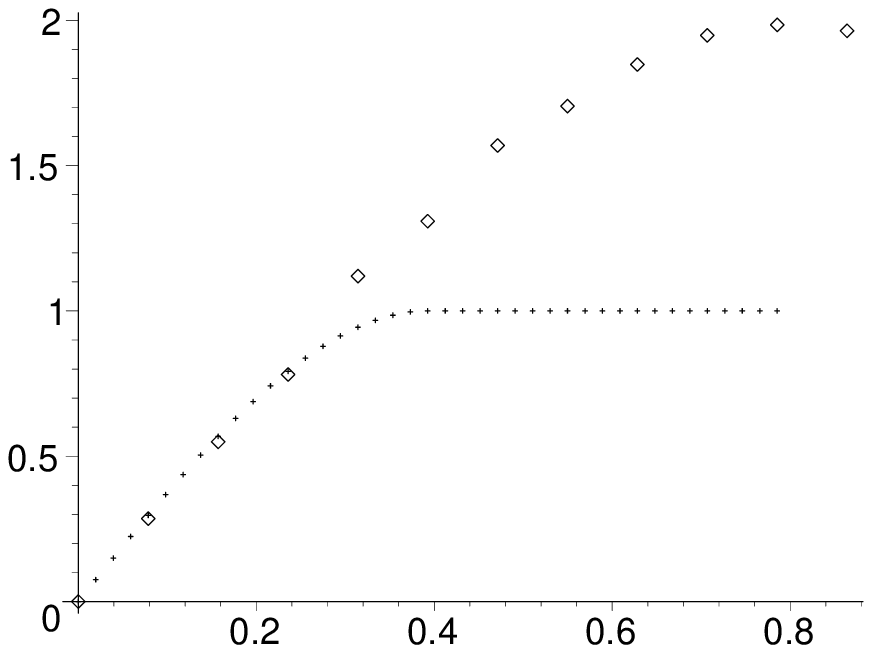}}
\put(0.5,0.02){$\alpha_1$}
\put(0.02,0.78){$\mathcal{EC}_{E}$}
\end{picture}
\caption{\label{graph:DCNOT}Single-copy entangling capacity for DCNOT family.  Crosses are for no ancillas and diamonds are for one ancilla on each side.}
\end{figure}

\begin{figure}
\setlength{\unitlength}{7cm}
\begin{picture}(1,0.85)
\put(0,0){\includegraphics[angle = 0, width = 7.0cm, height = 6.0cm]{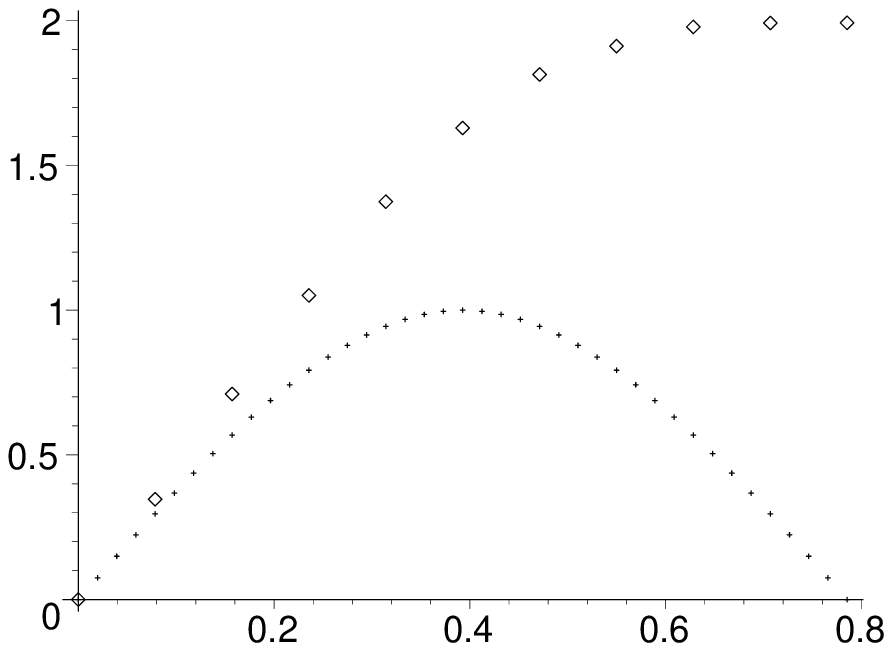}}
\put(0.5,0.02){$\alpha_1$}
\put(0.02,0.78){$\mathcal{EC}_{E}$}
\end{picture}
\caption{\label{graph:SWAP}Single-copy entangling capacity for SWAP family.  Crosses are for no ancillas and diamonds are for one ancilla on each side.}
\end{figure}

The simulations were run with both one and two ancillary qubits on each side.  Adding 1 ancillary qubit on each side increased the entangling capacity for the DCNOT and SWAP families (see Figs. \ref{graph:DCNOT} and \ref{graph:SWAP}), but there was no further increase on adding more ancillary qubits.  We conjecture that one ancillary qubit on each side is the most general system required to optimise single-copy entangling capacity.  Note that, for every $\alpha$, the SWAP family has a higher entangling capacity than the DCNOT family. This shows that the entangling capacity is generally a function of all three parameters ($\alpha_1, \alpha_2, \alpha_3$) of the unitary, in contrast to the case considered above where no ancillas are allowed.

\begin{figure}
\setlength{\unitlength}{7cm}
\begin{picture}(1,0.85)
\put(0,0){\includegraphics[angle = 0, width = 7.0cm, height = 6.0cm]{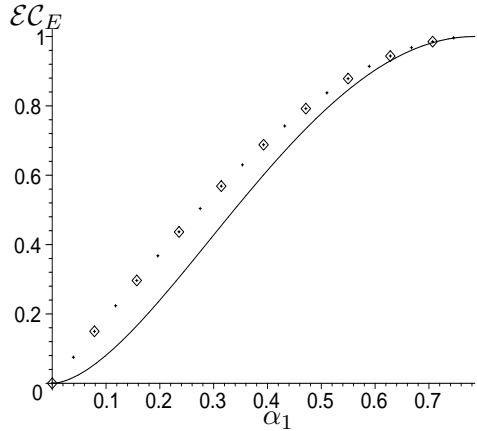}}
\put(0.5,0.02){$\alpha_1$}
\put(0.02,0.78){$\mathcal{EC}_{E}$}
\end{picture}
\caption{\label{graph:CNOT}Single-copy entangling capacity for CNOT family.  Crosses are for no ancillas, diamonds are for one ancilla on each side and the line shows the equivalent result when the starting state is restricted to be a product between Alice and Bob}
\end{figure}

For the CNOT family, adding ancillas had no effect at all (see FIG. \ref{graph:CNOT}).    In \cite{CiracDecomp}, the entangling capacity for the CNOT family starting from a product state with ancillas was found to be $H (\cos^2\alpha) = -\cos^2(\alpha)\log_2[\cos^2(\alpha)] - \sin^2(\alpha)\log_2[\sin^2(\alpha)]$.  No ancillas were required to achieve this capacity. Our results exceed this capacity, which demonstrates that allowing initial entanglement can still increase the entangling capacity even if ancillas are present.
\label{Anc}

\section{Collective Processing}
\label{Coll}

We now turn to the question of whether the entangling capacity is 
increased by applying $n$ copies of a unitary operation to pairs 
of qubits in the most general initial state which may be 
entangled and may contain ancillas.  The $n$-copy entangling capacity is then defined to be the optimal increase in entanglement over Alice and Bob's entire Hilbert space per application of the unitary.  We restrict our attention to 
the case where we have a pure state at every stage of the 
protocol, but note that the results also hold for the case where 
mixed states are allowed \cite{IBMent, IBMClass}.  In this 
setting, The unitaries may be applied simultaneously or one after 
another.  Collective LOCC may be performed on all the qubits 
between applications and each unitary may be applied to an 
arbitrarily chosen pairs of qubits.  However, all protocols of 
this form can be reduced to simpler protocols, which yield the 
same amount of entanglement.

First, observe that applying unitaries simultaneously is less 
general than applying them one after the other.  Second, because 
local unitary operations (e.g. local SWAP operations) can be 
applied as part of the LOCC, all the unitaries can be applied to 
the same pair of qubits.  Thus the problem reduces to a sequence 
of single-copy problems and we can do no better at each step than 
if we have the optimal single-copy entangled state available.

If initial entanglement is not available then collective 
processing can do better per use of the unitary, since we can 
make use of the first few copies of the unitary to generate 
entanglement, which can then be used to make a state with optimal 
initial entanglement.  This can then be used as the starting 
state for the subsequent copies.

Protocols that start with initial entanglement can outperform 
protocols that start with product states for all finite $n$.  
However, the asymptotic case, where $n \rightarrow \infty$, is 
more subtle because the operations of entanglement distillation 
and dilution are available for the states.  In the case where we 
start with product states, we can use up some of the first few 
operations to generate the entanglement required for the optimal 
initial state. Then we can keep diluting the entanglement of the 
states at each stage so that we always act on the best initial 
state.  The number of operations required for the first stage of 
this protocol is fixed and finite, so as $n \rightarrow \infty$ 
we will achieve the same entangling capacity as if we has started 
with initial entanglement.  This means that asymptotic entangling 
capacity of a unitary starting with a product state is the same 
as the capacity that would be obtained starting with initial 
entanglement.

\section{Conclusions}
We have shown that for all finite numbers of copies of $U_{AB}$, initial entanglement is required to achieve the optimal entangling capacity.  If this initial entanglement and ancillas are available, then collective processing does not help to achieve this maximum.

Our results have implications for the asymptotic interconvertibility of bipartite unitary operations.  For example, it is known that one can reversibly convert between a CNOT and a singlet state via LOCC.  Thus, one can asymptotically simulate the action of $n\mathcal{EC}_E(U_{AB})$ CNOTs using $n$ copies of $U_{AB}$ and LOCC by generating entanglement and then distilling or diluting it to singlets.  Further, it is impossible to generate more CNOTs than this, since otherwise one could generate more than $\mathcal{EC}_E(U_{AB})$ e-bits per application of $U_{AB}$ by first converting to CNOTs and then using them to generate singlet states.  More generally, it is not known whether converting between any unitary operation and entanglement via LOCC is reversible (i.e. whether one can asymptotically generate $n$ copies of $U_{AB}$ acting on an arbitrary input state given $n\mathcal{EC}_E(U_{AB})$ e-bits).  However, $n\mathcal{EC}_E(U_{AB})$ is a lower bound on how much entanglement is needed to generate $n$ copies of $U_{AB}$.  Also, $\frac{\mathcal{EC}_E(U_1)}{\mathcal{EC}_E(U_2)}$ is an upper bound on how many copies of a bipartite unitary $U_2$ can be generated asymptotically per application of another bipartite unitary $U_1$.  Whether these bounds can be achieved remains an open question.

\label{Conc}

\begin{acknowledgments}
We are very grateful to C. H. Bennett, D. Leung and S. Popescu for many stimulating discussions about non-local properties of unitaries and to C. H. Bennett, A. Harrow, D. Leung and J. Smolin for describing their recent results to us prior to their publication.  We gratefully acknowledge funding from the European Union under the project EQUIP (contract IST-1999-11063).
\end{acknowledgments}


\begin{thebibliography}{30}
\expandafter\ifx\csname natexlab\endcsname\relax\def\natexlab#1{#1}\fi
\expandafter\ifx\csname bibnamefont\endcsname\relax
  \def\bibnamefont#1{#1}\fi
\expandafter\ifx\csname bibfnamefont\endcsname\relax
  \def\bibfnamefont#1{#1}\fi
\expandafter\ifx\csname citenamefont\endcsname\relax
  \def\citenamefont#1{#1}\fi
\expandafter\ifx\csname url\endcsname\relax
  \def\url#1{\texttt{#1}}\fi
\expandafter\ifx\csname urlprefix\endcsname\relax\def\urlprefix{URL }\fi
\providecommand{\bibinfo}[2]{#2}
\providecommand{\eprint}[2][]{\url{#2}}

\bibitem[{\citenamefont{Nielsen}(1999)}]{NeilMon}
\bibinfo{author}{\bibfnamefont{M.~A.} \bibnamefont{Nielsen}},
  \bibinfo{journal}{Phys. Rev. Lett.}
  \textbf{\bibinfo{volume}{83}}(\bibinfo{number}{2}), \bibinfo{pages}{436}
  (\bibinfo{year}{1999}).

\bibitem[{\citenamefont{Linden et~al.}(1998)\citenamefont{Linden, Massar, and
  Popescu}}]{PurColl}
\bibinfo{author}{\bibfnamefont{N.}~\bibnamefont{Linden}},
  \bibinfo{author}{\bibfnamefont{S.}~\bibnamefont{Massar}}, \bibnamefont{and}
  \bibinfo{author}{\bibfnamefont{S.}~\bibnamefont{Popescu}},
  \bibinfo{journal}{Phys. Rev. Lett.} \textbf{\bibinfo{volume}{81}},
  \bibinfo{pages}{3279} (\bibinfo{year}{1998}), \eprint{quant-ph/9805001}.

\bibitem[{\citenamefont{Bennett et~al.}(1996)\citenamefont{Bennett, Bernstein,
  Popescu, and Schumacher}}]{Concentrate}
\bibinfo{author}{\bibfnamefont{C.~H.} \bibnamefont{Bennett}},
  \bibinfo{author}{\bibfnamefont{H.~J.} \bibnamefont{Bernstein}},
  \bibinfo{author}{\bibfnamefont{S.}~\bibnamefont{Popescu}}, \bibnamefont{and}
  \bibinfo{author}{\bibfnamefont{B.}~\bibnamefont{Schumacher}},
  \bibinfo{journal}{Phys. Rev. A} \textbf{\bibinfo{volume}{54}},
  \bibinfo{pages}{4707} (\bibinfo{year}{1996}), \eprint{quant-ph/9511030}.

\bibitem[{\citenamefont{Linden et~al.}(1999)\citenamefont{Linden, Popescu,
  Schumacher, and Westmoreland}}]{Reverse}
\bibinfo{author}{\bibfnamefont{N.}~\bibnamefont{Linden}},
  \bibinfo{author}{\bibfnamefont{S.}~\bibnamefont{Popescu}},
  \bibinfo{author}{\bibfnamefont{B.}~\bibnamefont{Schumacher}},
  \bibnamefont{and}
  \bibinfo{author}{\bibfnamefont{M.}~\bibnamefont{Westmoreland}},
  \emph{\bibinfo{title}{Reversibility of local transformations of multiparticle
  entanglement}} (\bibinfo{year}{1999}), \eprint{quant-ph/9912039}.

\bibitem[{\citenamefont{Bennett et~al.}(2001)\citenamefont{Bennett, Cirac,
  Leifer, Leung, Linden, Popescu, and Vidal}}]{2qSim}
\bibinfo{author}{\bibfnamefont{C.~H.} \bibnamefont{Bennett}},
  \bibinfo{author}{\bibfnamefont{J.~I.} \bibnamefont{Cirac}},
  \bibinfo{author}{\bibfnamefont{M.~S.} \bibnamefont{Leifer}},
  \bibinfo{author}{\bibfnamefont{D.~W.} \bibnamefont{Leung}},
  \bibinfo{author}{\bibfnamefont{N.}~\bibnamefont{Linden}},
  \bibinfo{author}{\bibfnamefont{S.}~\bibnamefont{Popescu}}, \bibnamefont{and}
  \bibinfo{author}{\bibfnamefont{G.}~\bibnamefont{Vidal}},
  \emph{\bibinfo{title}{Optimal simulation of two-qubit hamiltonians using
  general local operations}} (\bibinfo{year}{2001}), \eprint{quant-ph/0107035}.

\bibitem[{\citenamefont{Khaneja et~al.}(2001)\citenamefont{Khaneja, Brockett,
  and Glaser}}]{UniSim}
\bibinfo{author}{\bibfnamefont{N.}~\bibnamefont{Khaneja}},
  \bibinfo{author}{\bibfnamefont{R.}~\bibnamefont{Brockett}}, \bibnamefont{and}
  \bibinfo{author}{\bibfnamefont{S.~J.} \bibnamefont{Glaser}},
  \bibinfo{journal}{Phys. Rev. A} \textbf{\bibinfo{volume}{63}},
  \bibinfo{pages}{032308 1} (\bibinfo{year}{2001}).

\bibitem[{\citenamefont{Dodd et~al.}(2001)\citenamefont{Dodd, Nielsen, Bremner,
  and Thew}}]{NeilHam1}
\bibinfo{author}{\bibfnamefont{J.}~\bibnamefont{Dodd}},
  \bibinfo{author}{\bibfnamefont{M.}~\bibnamefont{Nielsen}},
  \bibinfo{author}{\bibfnamefont{M.}~\bibnamefont{Bremner}}, \bibnamefont{and}
  \bibinfo{author}{\bibfnamefont{R.}~\bibnamefont{Thew}},
  \emph{\bibinfo{title}{Universal quantum computation and simulation using any
  entangling hamiltonian and local unitaries}} (\bibinfo{year}{2001}),
  \eprint{quant-ph/0106064}.

\bibitem[{\citenamefont{Nielsen et~al.}(2001)\citenamefont{Nielsen, Bremner,
  Dodd, Childs, and Dawson}}]{NeilHam2}
\bibinfo{author}{\bibfnamefont{M.}~\bibnamefont{Nielsen}},
  \bibinfo{author}{\bibfnamefont{M.}~\bibnamefont{Bremner}},
  \bibinfo{author}{\bibfnamefont{J.}~\bibnamefont{Dodd}},
  \bibinfo{author}{\bibfnamefont{A.}~\bibnamefont{Childs}}, \bibnamefont{and}
  \bibinfo{author}{\bibfnamefont{C.}~\bibnamefont{Dawson}},
  \emph{\bibinfo{title}{Universal simulation of hamiltonian dynamics for
  qdits}} (\bibinfo{year}{2001}), \eprint{quant-ph/0109064}.

\bibitem[{\citenamefont{Wocjan et~al.}(2001{\natexlab{a}})\citenamefont{Wocjan,
  Janzing, and Beth}}]{Beth1}
\bibinfo{author}{\bibfnamefont{P.}~\bibnamefont{Wocjan}},
  \bibinfo{author}{\bibfnamefont{D.}~\bibnamefont{Janzing}}, \bibnamefont{and}
  \bibinfo{author}{\bibfnamefont{T.}~\bibnamefont{Beth}},
  \emph{\bibinfo{title}{Simulating arbitrary pair-interactions by a given
  hamiltonian}} (\bibinfo{year}{2001}{\natexlab{a}}),
  \eprint{quant-ph/0106077}.

\bibitem[{\citenamefont{Janzing et~al.}(2001)\citenamefont{Janzing, Wocjan, and
  Beth}}]{Beth2}
\bibinfo{author}{\bibfnamefont{D.}~\bibnamefont{Janzing}},
  \bibinfo{author}{\bibfnamefont{P.}~\bibnamefont{Wocjan}}, \bibnamefont{and}
  \bibinfo{author}{\bibfnamefont{T.}~\bibnamefont{Beth}},
  \emph{\bibinfo{title}{Complexity of decoupling and time-reversal for n spins
  with pair-interactions}} (\bibinfo{year}{2001}), \eprint{quant-ph/0106085}.

\bibitem[{\citenamefont{Wocjan et~al.}(2001{\natexlab{b}})\citenamefont{Wocjan,
  Roetteler, Janzing, and Beth}}]{Beth3}
\bibinfo{author}{\bibfnamefont{P.}~\bibnamefont{Wocjan}},
  \bibinfo{author}{\bibfnamefont{M.}~\bibnamefont{Roetteler}},
  \bibinfo{author}{\bibfnamefont{D.}~\bibnamefont{Janzing}}, \bibnamefont{and}
  \bibinfo{author}{\bibfnamefont{T.}~\bibnamefont{Beth}},
  \emph{\bibinfo{title}{Universal simulation of hamiltonians using a finite set
  of control operations}} (\bibinfo{year}{2001}{\natexlab{b}}),
  \eprint{quant-ph/0109063}.

\bibitem[{\citenamefont{Wocjan et~al.}(2001{\natexlab{c}})\citenamefont{Wocjan,
  Roetteler, Janzing, and Beth}}]{Beth4}
\bibinfo{author}{\bibfnamefont{P.}~\bibnamefont{Wocjan}},
  \bibinfo{author}{\bibfnamefont{M.}~\bibnamefont{Roetteler}},
  \bibinfo{author}{\bibfnamefont{D.}~\bibnamefont{Janzing}}, \bibnamefont{and}
  \bibinfo{author}{\bibfnamefont{T.}~\bibnamefont{Beth}},
  \emph{\bibinfo{title}{Simulating hamiltonians in quantum networks}}
  (\bibinfo{year}{2001}{\natexlab{c}}), \eprint{quant-ph/0109088}.

\bibitem[{\citenamefont{Leung}(2001)}]{Leung1}
\bibinfo{author}{\bibfnamefont{D.}~\bibnamefont{Leung}},
  \emph{\bibinfo{title}{Simulation and reversal of n-qubit hamiltonians using
  hadamard matrices}} (\bibinfo{year}{2001}), \eprint{quant-ph/0107041}.

\bibitem[{\citenamefont{Chen}(2001)}]{ChenSim}
\bibinfo{author}{\bibfnamefont{H.}~\bibnamefont{Chen}},
  \emph{\bibinfo{title}{Necessary conditions for efficient simulation of
  hamiltonians using local unitary transformations}} (\bibinfo{year}{2001}),
  \eprint{quant-ph/0109115}.

\bibitem[{\citenamefont{Vidal and Cirac}(2001{\natexlab{a}})}]{CiracSim1}
\bibinfo{author}{\bibfnamefont{G.}~\bibnamefont{Vidal}} \bibnamefont{and}
  \bibinfo{author}{\bibfnamefont{J.}~\bibnamefont{Cirac}},
  \emph{\bibinfo{title}{Optimal simulation of nonlocal hamiltonians using local
  operations and classical communication}}
  (\bibinfo{year}{2001}{\natexlab{a}}), \eprint{quant-ph/0108076}.

\bibitem[{\citenamefont{Vidal and Cirac}(2001{\natexlab{b}})}]{CiracSim2}
\bibinfo{author}{\bibfnamefont{G.}~\bibnamefont{Vidal}} \bibnamefont{and}
  \bibinfo{author}{\bibfnamefont{J.}~\bibnamefont{Cirac}},
  \emph{\bibinfo{title}{Catalysis in non-local quantum operations}}
  (\bibinfo{year}{2001}{\natexlab{b}}), \eprint{quant-ph/0108077}.

\bibitem[{\citenamefont{Vidal et~al.}(2001)\citenamefont{Vidal, Hammerer, and
  Cirac}}]{CiracSim3}
\bibinfo{author}{\bibfnamefont{G.}~\bibnamefont{Vidal}},
  \bibinfo{author}{\bibfnamefont{K.}~\bibnamefont{Hammerer}}, \bibnamefont{and}
  \bibinfo{author}{\bibfnamefont{J.}~\bibnamefont{Cirac}},
  \emph{\bibinfo{title}{Interaction cost of non-local gates}}
  (\bibinfo{year}{2001}), \eprint{quant-ph/0112168}.

\bibitem[{\citenamefont{D{\"u}r
  et~al.}(2001{\natexlab{a}})\citenamefont{D{\"u}r, Vidal, and
  Cirac}}]{CiracCon1}
\bibinfo{author}{\bibfnamefont{W.}~\bibnamefont{D{\"u}r}},
  \bibinfo{author}{\bibfnamefont{G.}~\bibnamefont{Vidal}}, \bibnamefont{and}
  \bibinfo{author}{\bibfnamefont{J.}~\bibnamefont{Cirac}},
  \emph{\bibinfo{title}{Optimal conversion of non-local unitary operations}}
  (\bibinfo{year}{2001}{\natexlab{a}}), \eprint{quant-ph/0112124}.

\bibitem[{\citenamefont{D{\"u}r and Cirac}(2001)}]{CiracCon2}
\bibinfo{author}{\bibfnamefont{W.}~\bibnamefont{D{\"u}r}} \bibnamefont{and}
  \bibinfo{author}{\bibfnamefont{J.}~\bibnamefont{Cirac}},
  \emph{\bibinfo{title}{Equivalence classes of non-local unitary operations}}
  (\bibinfo{year}{2001}), \eprint{quant-ph/0201112}.

\bibitem[{\citenamefont{D{\"u}r
  et~al.}(2001{\natexlab{b}})\citenamefont{D{\"u}r, Vidal, Cirac, Linden, and
  Popescu}}]{Hamil1}
\bibinfo{author}{\bibfnamefont{W.}~\bibnamefont{D{\"u}r}},
  \bibinfo{author}{\bibfnamefont{G.}~\bibnamefont{Vidal}},
  \bibinfo{author}{\bibfnamefont{J.}~\bibnamefont{Cirac}},
  \bibinfo{author}{\bibfnamefont{N.}~\bibnamefont{Linden}}, \bibnamefont{and}
  \bibinfo{author}{\bibfnamefont{S.}~\bibnamefont{Popescu}},
  \bibinfo{journal}{Phys. Rev. Lett.} \textbf{\bibinfo{volume}{87}},
  \bibinfo{pages}{137901} (\bibinfo{year}{2001}{\natexlab{b}}),
  \bibinfo{note}{to appear in Phys. Rev. Lett.}, \eprint{quant-ph/0006034}.

\bibitem[{\citenamefont{Kraus and Cirac}(2001)}]{CiracDecomp}
\bibinfo{author}{\bibfnamefont{B.}~\bibnamefont{Kraus}} \bibnamefont{and}
  \bibinfo{author}{\bibfnamefont{J.}~\bibnamefont{Cirac}},
  \bibinfo{journal}{Phys. Rev. A} \textbf{\bibinfo{volume}{63}},
  \bibinfo{pages}{062309} (\bibinfo{year}{2001}), \eprint{quant-ph/0011050}.

\bibitem[{\citenamefont{Zanardi et~al.}(2000)\citenamefont{Zanardi, Zalka, and
  Faoro}}]{Zanardi1}
\bibinfo{author}{\bibfnamefont{P.}~\bibnamefont{Zanardi}},
  \bibinfo{author}{\bibfnamefont{C.}~\bibnamefont{Zalka}}, \bibnamefont{and}
  \bibinfo{author}{\bibfnamefont{L.}~\bibnamefont{Faoro}},
  \bibinfo{journal}{Physical Review A} \textbf{\bibinfo{volume}{62}},
  \bibinfo{pages}{030301} (\bibinfo{year}{2000}).

\bibitem[{\citenamefont{Zanardi}(2001)}]{Zanardi2}
\bibinfo{author}{\bibfnamefont{P.}~\bibnamefont{Zanardi}},
  \bibinfo{journal}{Phys. Rev. A} \textbf{\bibinfo{volume}{63}},
  \bibinfo{pages}{040304(R)} (\bibinfo{year}{2001}), \eprint{quant-ph/0010074}.

\bibitem[{\citenamefont{Cirac et~al.}(2001)\citenamefont{Cirac, D{\"u}r, Kraus,
  and Lewenstein}}]{CiracStatOp}
\bibinfo{author}{\bibfnamefont{J.}~\bibnamefont{Cirac}},
  \bibinfo{author}{\bibfnamefont{W.}~\bibnamefont{D{\"u}r}},
  \bibinfo{author}{\bibfnamefont{B.}~\bibnamefont{Kraus}}, \bibnamefont{and}
  \bibinfo{author}{\bibfnamefont{M.}~\bibnamefont{Lewenstein}},
  \bibinfo{journal}{Phys. Rev. Lett.} \textbf{\bibinfo{volume}{86}},
  \bibinfo{pages}{544} (\bibinfo{year}{2001}), \eprint{quant-ph/0007057}.

\bibitem[{\citenamefont{Collins et~al.}(2001)\citenamefont{Collins, Linden, and
  Popescu}}]{NonOp}
\bibinfo{author}{\bibfnamefont{D.}~\bibnamefont{Collins}},
  \bibinfo{author}{\bibfnamefont{N.}~\bibnamefont{Linden}}, \bibnamefont{and}
  \bibinfo{author}{\bibfnamefont{S.}~\bibnamefont{Popescu}},
  \bibinfo{journal}{Phys. Rev. A} \textbf{\bibinfo{volume}{64}},
  \bibinfo{pages}{032302} (\bibinfo{year}{2001}), \eprint{quant-ph/0005102}.

\bibitem[{\citenamefont{Eisert et~al.}(2000)\citenamefont{Eisert, Jacobs,
  Papadopoulos, and Plenio}}]{PlenOp}
\bibinfo{author}{\bibfnamefont{J.}~\bibnamefont{Eisert}},
  \bibinfo{author}{\bibfnamefont{K.~A.} \bibnamefont{Jacobs}},
  \bibinfo{author}{\bibfnamefont{P.}~\bibnamefont{Papadopoulos}},
  \bibnamefont{and} \bibinfo{author}{\bibfnamefont{M.~B.}
  \bibnamefont{Plenio}}, \bibinfo{journal}{Phys. Rev. A}
  \textbf{\bibinfo{volume}{62}}, \bibinfo{pages}{052317}
  (\bibinfo{year}{2000}), \eprint{quant-ph/0005101}.

\bibitem[{\citenamefont{Leung and Bennett}(2002)}]{IBMent}
\bibinfo{author}{\bibfnamefont{D.}~\bibnamefont{Leung}} \bibnamefont{and}
  \bibinfo{author}{\bibfnamefont{C.~H.} \bibnamefont{Bennett}},
  \bibinfo{howpublished}{private communication} (\bibinfo{year}{2002}).

\bibitem[{\citenamefont{Bennett et~al.}(2002)\citenamefont{Bennett, Harrow,
  Leung, and Smolin}}]{IBMClass}
\bibinfo{author}{\bibfnamefont{C.~H.} \bibnamefont{Bennett}},
  \bibinfo{author}{\bibfnamefont{A.}~\bibnamefont{Harrow}},
  \bibinfo{author}{\bibfnamefont{D.~W.} \bibnamefont{Leung}}, \bibnamefont{and}
  \bibinfo{author}{\bibfnamefont{J.~A.} \bibnamefont{Smolin}},
  \emph{\bibinfo{title}{On the capacities of bipartite hamiltonians and unitary
  gates}} (\bibinfo{year}{2002}), \eprint{quant-ph/0205057}.

\bibitem[{\citenamefont{Makhlin}(2000)}]{Makhlin}
\bibinfo{author}{\bibfnamefont{Y.}~\bibnamefont{Makhlin}},
  \emph{\bibinfo{title}{Nonlocal properties of two-qubit gates and mixed states
  and optimization of quantum computations}} (\bibinfo{year}{2000}),
  \eprint{quant-ph/0002045}.

\bibitem[{\citenamefont{Wootters}(1998)}]{WootCon}
\bibinfo{author}{\bibfnamefont{W.~K.} \bibnamefont{Wootters}},
  \bibinfo{journal}{Phys. Rev. Lett.} \textbf{\bibinfo{volume}{80}},
  \bibinfo{pages}{2245} (\bibinfo{year}{1998}).

\end{thebibliography}
\end{document}